\documentclass[conference]{IEEEtran}
\IEEEoverridecommandlockouts
\usepackage{cite}
\usepackage{amsmath,amssymb,amsfonts}
\usepackage{algorithmic}
\usepackage{graphicx}
\usepackage{textcomp}
\def\BibTeX{{\rm B\kern-.05em{\sc i\kern-.025em b}\kern-.08em
    T\kern-.1667em\lower.7ex\hbox{E}\kern-.125emX}}

\usepackage{times}
\usepackage{soul}
\usepackage{url}
\usepackage[hidelinks]{hyperref}
\usepackage[utf8]{inputenc}
\usepackage{algorithm}
\usepackage[switch]{lineno}
\usepackage[small]{caption}
\usepackage{kotex}
\usepackage{multirow}
\usepackage{multicol}
\usepackage[table]{xcolor}
\usepackage{tabularx}
\usepackage{booktabs}
\usepackage{makecell}

\newcommand{\kw}[1]{\colorbox{yellow!35}{\strut #1}}

\newcommand{\jh}[1]{\colorbox{cyan!20}{\strut #1}}

\newcommand{\ok}[1]{\textcolor{blue!75!black}{\textbf{#1}}}
\newcommand{\bad}[1]{\textcolor{red!75!black}{\textbf{#1}}}

\begin{document}

\title{TRAE: Trustworthy Automated ESG Annotation via Multi-Agent Consensus and Confidence Assessment}

\author{
\IEEEauthorblockN{Jaeyoung Lee}
\IEEEauthorblockA{
\textit{Department of Data Science} \\
\textit{Hanyang University}\\
Seoul, Republic of Korea \\
jaeylee@hanyang.ac.kr
}
\and
\IEEEauthorblockN{Misuk Kim\textsuperscript{*}}
\IEEEauthorblockA{
\textit{Department of Data Science} \\
\textit{Department of Artificial Intelligence} \\
\textit{Hanyang University}\\
Seoul, Republic of Korea \\
misukkim@hanyang.ac.kr\\
}
\thanks{\textsuperscript{*}Corresponding author.}
}

\maketitle

\begin{abstract}
With the growing demand for ESG disclosure and evaluation, constructing high-quality datasets for automatically extracting ESG-related information from large-scale documents has become essential. However, ESG annotation remains challenging due to strong context dependence, ambiguous label boundaries, and the need for domain expertise, resulting in high annotation cost and limited scalability. To address these challenges, we propose \textsc{TRAE} (\textbf{TR}ustworthy \textbf{A}utomated \textbf{E}SG annotation), a multi-agent framework for scalable and reliable ESG data construction. Unlike direct LLM-based annotation, \textsc{TRAE} evaluates annotation reliability by modeling agreement, confidence, and reasoning consistency among multiple agents through a structured multi-step reasoning process. Experimental results demonstrate that \textsc{TRAE} improves annotation performance compared with existing automated annotation strategies. Furthermore, \textsc{TRAE} supports a hybrid human-AI annotation strategy that assigns expert labeling only to uncertain samples, reducing annotation cost while preserving data quality. Using \textsc{TRAE}, we construct a high-quality ESG dataset containing approximately 20,000 instances with complete reasoning traces, intermediate agent outputs, and reliability information to enhance transparency and reproducibility. Overall, this work provides a practical and scalable framework for trustworthy ESG data construction. The code and data are publicly available at: \url{https://github.com/nowzer0/TRAE}.

\end{abstract}


\section{Introduction}
To evaluate sustainability as a key criterion for assessing corporate activities, Environmental, Social, and Governance (ESG) aspects are widely considered. Companies disclose, in various forms, environmental impacts (e.g., climate, resources, pollution), social issues (e.g., labor, human rights, supply chains), and governance factors (e.g., decision-making structures, risk management, disclosure reliability). Stakeholders use this information to assess sustainability management and risks and to inform policy and investment decisions. However, ESG information is dispersed across large volumes of unstructured documents (e.g., news, sustainability reports, business reports), imposing substantial costs to obtain and analyze. Therefore, the importance of technologies that automatically extract and classify ESG-related information from documents is increasing~\cite{lee2023esg}.
ESG information extraction is a prerequisite for downstream tasks such as evaluation, risk detection, monitoring, and model training, and its quality strongly affects overall performance and reliability. However, ESG information extraction is constrained by the lack of high-quality labeled ESG data. Because ESG interpretations are context-dependent with ambiguous label boundaries, and accurate judgment requires domain expertise, expert-centered manual annotation is costly and time-consuming~\cite{he2025ESGeni}. As a result, building reliable large-scale ESG datasets is difficult, limiting the identification of clear ESG-related information.

Recent advances in large language models (LLMs) suggest that annotation processes can be automated given their natural language understanding and reasoning capabilities~\cite{zhao2023survey}. Inference techniques that mitigate response variability via multi-sampling or improve judgment consistency through cross-verification among multiple agents further support more effective automation~\cite{wang2023selfconsistency,chen2024reconcile}. However, in highly context-dependent domains with fine-grained label boundaries such as ESG, a single inference outcome can be skewed by inherent LLM biases, and annotation errors with unclear justification may accumulate. Therefore, clear reliability criteria and an operational strategy that determines when to apply additional reasoning or involve experts—while considering the cost--quality trade-off—are needed. Moreover, annotation outputs should be released with the decision process and supporting rationale to ensure transparency and reproducibility.

To this end, we propose \textsc{TRAE} (\textbf{TR}ustworthy \textbf{A}utomated \textbf{E}SG annotation), a multi-agent framework for trustworthy automated ESG annotation. \textsc{TRAE} collects candidates from multiple LLM agents through multi-stage reasoning and evaluates stage-wise outcomes using inter-agent consensus and confidence. It then applies conditional routing: reliable samples are automatically annotated, while uncertain samples are sent to additional reasoning or expert annotation. This expands LLM-based automation while adding verification in error-prone regions, enabling efficient and trustworthy annotation that balances cost and quality. We validate \textsc{TRAE} by measuring agreement with expert annotations and downstream task performance. Results show high expert agreement, and ESG classifiers trained on automatically annotated data achieve significant performance improvements. We further show that a hybrid strategy combining automated and expert annotation reduces cost while still enabling reliable, high-quality dataset construction. Our contributions are as follows.
\begin{itemize}
\item We propose \textsc{TRAE}, a multi-agent framework for trustworthy ESG annotation that combines multi-stage reasoning with consensus- and confidence-based conditional routing.
\item We validate \textsc{TRAE} via agreement with expert annotations and downstream experiments, showing that an expert-in-the-loop hybrid strategy reduces annotation cost while enabling reliable, high-quality dataset construction.
\item In collaboration with ESG experts, we build an ESG information extraction dataset of approximately 20,000 instances using \textsc{TRAE}, and improve transparency and reproducibility by releasing stage-wise reasoning paths and intermediate outputs.
\end{itemize}
Overall, \textsc{TRAE} enables efficient and trustworthy annotation, alleviating the high cost and limited scalability of ESG data construction and supporting downstream ESG tasks, including ESG information extraction.

\section{Related Works}
\subsection{LLMs as Annotators}
Using LLMs as annotators to automatically generate labels is promising in terms of time and cost, and many studies have adopted this approach~\cite{veselovsky2023artificial,Gilardi2023ChatGPT,Chiang2023CanLL}. However, annotation performance varies substantially across tasks: it reaches human-level performance in some cases~\cite{zhu2023chatgpt,tornberg2023chatgpt} but remains limited in others~\cite{reiss2023testing,ziems2024large}. Accordingly, recent work increasingly evaluates LLM-based annotation against verifiable criteria~\cite{ni2024afacta}. To stabilize annotation, some studies strengthen prompts or incorporate external knowledge~\cite{he2024annollm,ding-2023-gpt}, while others argue that human intervention is necessary~\cite{gligoric2025unconfident,kim2024meganno}. Beyond NLP, LLM-based annotation has also been discussed in text-as-data social science research~\cite{Bali2024Cam}; nevertheless, more work is needed on operational approaches that ensure both domain trustworthiness and quality.

\subsection{LLMs for ESG Data Construction}
Recent studies have addressed ESG data scarcity, a major bottleneck in ESG research~\cite{lee2024esg}. However, ESG dataset construction remains time- and cost-intensive, motivating LLM-based approaches. Prior work builds ESG datasets by combining keyword retrieval with LLM API labeling and training models on the resulting data~\cite{xia2024using}. Others incorporate GRI index mapping, embedding-based similarity, and LLM filtering into a refinement pipeline to remove noisy labels~\cite{mousavian2025automatic}, and further improvements have been reported via GPT-based multi-label annotation with imbalance correction~\cite{Wu2024Enhancing}. Expert involvement has also been explored to strengthen trustworthiness: ESGenius generates candidate questions grounded in standards such as GRI and SASB using LLMs, which ESG experts then verify and revise to build an evaluation benchmark~\cite{he2025ESGeni}.

Existing LLM-based ESG data construction studies rely on automatically generated annotations but provide limited verification or reporting of label-policy trustworthiness and reasoning rationales. Given ESG’s context dependence and ambiguous label boundaries, automated annotation alone can accumulate errors and constrain downstream use. Therefore, a trustworthy procedure is needed that clearly separates reliability from uncertainty and ensures quality via additional verification for uncertain cases.
\begin{figure*}[h!]
    \centering
    \includegraphics[width=1.0\linewidth]{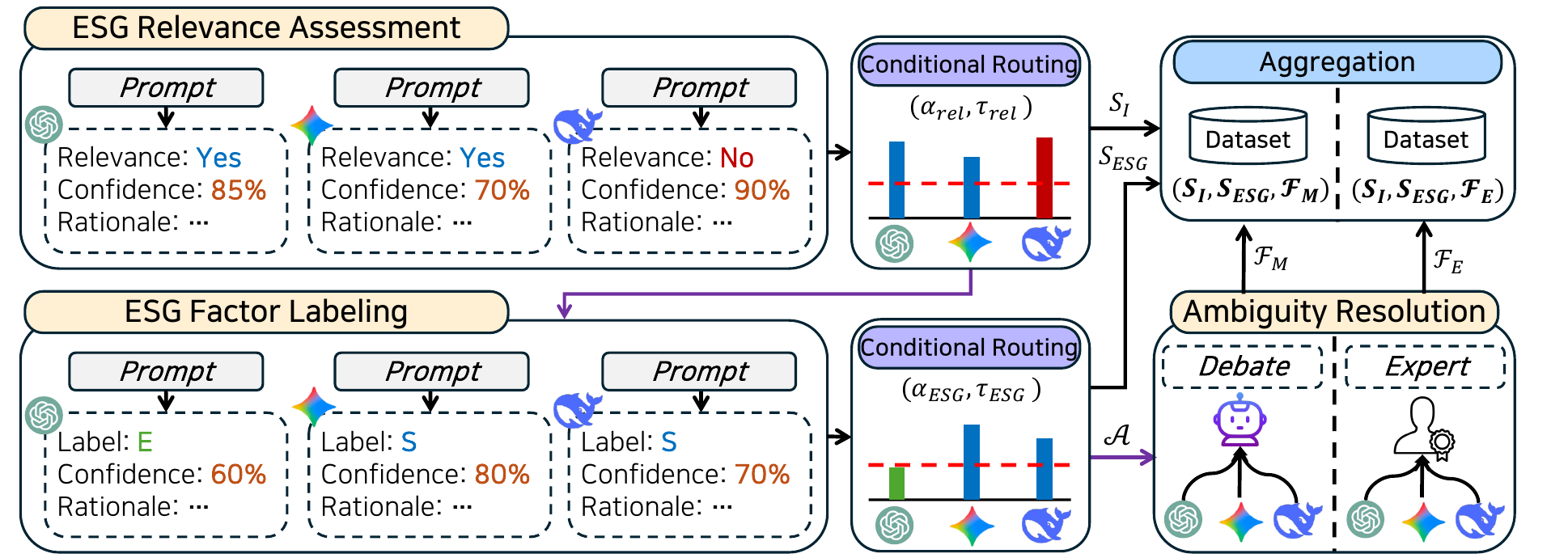}
    \caption{Overview of \textsc{TRAE}, a multi-agent ESG annotation framework composed of three modules—Relevance Assessment, ESG Factor Labeling, and Ambiguity Resolution—connected via consensus- and confidence-based conditional routing to finalize reliable labels and route uncertain samples to additional verification.}
    \label{fig2:overview}
\end{figure*}

\section{\textsc{TRAE}}
\textsc{TRAE} is composed of heterogeneous LLM agents to mitigate error accumulation from single-LLM biases~\cite{chen2024reconcile} and consists of three modules (Sections~\ref{module1}$\sim$\ref{module3}): (i) Relevance Assessment to determine whether information is ESG-related, (ii) ESG Factor Labeling to identify the specific ESG factor, and (iii) Ambiguity Resolution to finalize ambiguous samples via additional reasoning or verification (Fig.~\ref{fig2:overview}). Within each module, \textsc{TRAE} applies conditional routing based on inter-agent consensus and confidence to determine the processing path: reliable samples are finalized early via automated annotation, while uncertain samples are forwarded to subsequent modules or routed to final verification. Finally, \textsc{TRAE} aggregates stage-wise outcomes to construct the dataset and records each sample’s processing path and supporting rationale to enhance transparency and reproducibility.

\textit{\textbf{Prompt.}} We use Chain-of-Thought~(CoT)-based prompts in all modules~\cite{wei2022chain}, and assign each agent a role—such as an ESG expert, a reviewer, or a Judge responsible for the final decision—via the system prompt. Each agent’s response is designed to include an annotation, a confidence score, and a decision rationale. To prevent confidence scores from being produced in arbitrary ranges and thus becoming less comparable, we provide explicit instructions for a fixed confidence scale. In addition, to ensure labeling reliability and consistency of criteria, we curate and unify ESG issues presented by major ESG frameworks and standard-setters—MSCI~(Morgan Stanley Capital International), SASB~(Sustainability Accounting Standards Board), the GRI Standards~(Global Reporting Initiative), and TCFD~(Task Force on Climate-related Financial Disclosures)—through domain experts, and use them as annotation guidelines.

\textit{\textbf{Conditional Routing.}} \textsc{TRAE} applies conditional routing to determine each sample’s processing path based on multi-agent responses in each module. In each module, $K$ agents produce an annotation outcome, a confidence score, and a rationale; routing jointly evaluates inter-agent consensus and confidence. Consensus indicates whether agents output the same annotation, while confidence is each agent’s self-reported certainty and serves as a threshold to control automatic annotation. Each module defines a consensus threshold $\alpha_0$ and a confidence threshold $\tau_0$, and automatically annotates only samples that satisfy both. Thus, a label is finalized only when consensus and confidence are sufficiently high; otherwise, the sample is forwarded to the next module or separated into an uncertain set. This design expands automation while mitigating error accumulation in uncertain regions. We treat $\alpha_0$ and $\tau_0$ as hyperparameters and empirically study the automation--quality trade-off.

\subsection{Relevance Assessment}
\label{module1}
The Relevance Assessment module determines whether a sentence is ESG-related. Sentences from corporate disclosures, reports, and news often include substantial non-ESG content (e.g., general financial information, company introductions, and marketing phrases). This module explicitly filters such non-ESG sentences in advance, reducing unnecessary processing and noise in later stages. Agents assess each sentence’s ESG relevance and output the label with a confidence score and supporting rationale. Via conditional routing, reliably finalized samples $\mathcal{S}_{I}$ are labeled as \texttt{I(Irrelevant)} and excluded, while the remaining samples are forwarded to the ESG Factor Labeling module.

\subsection{ESG Factor Labeling}
\label{module2}
The ESG Factor Labeling module identifies the ESG factor for potential ESG candidates while re-validating ESG relevance. As the core annotation step, each sample is labeled according to the fine-grained factors in the unified ESG guidelines. Agents assign \texttt{E} (Environmental), \texttt{S} (Social), \texttt{G} (Governance), or \texttt{I}, and output the label with a confidence score and supporting rationale. Via conditional routing, reliably finalized samples $\mathcal{S}_{ESG}$ are annotated into four classes (\texttt{E, S, G, I}), while samples that fail the routing criteria are placed in the uncertain set $\mathcal{A}$ and forwarded to the Ambiguity Resolution module.

\subsection{Ambiguity Resolution}
\label{module3}
The Ambiguity Resolution module finalizes annotations for the uncertain set $\mathcal{A}$ that does not satisfy the conditional routing criteria. Depending on the setting, it follows either an expert-in-the-loop or a fully automated path to mitigate error accumulation while ensuring trustworthy final labels. In the fully automated setting, multiple agents generate claims supporting their initial predictions, and a higher-capacity Judge agent synthesizes the initial outputs and claims to determine the final label via a debate-style strategy. In the expert-in-the-loop setting, a domain expert finalizes the label using the annotation guidelines and model-generated candidate labels and rationales as reference information to improve efficiency. Accordingly, samples in $\mathcal{A}$ are labeled as either the fully automated set $\mathcal{F}_{M}$ or the expert-involved set $\mathcal{F}_{E}$.

Finally, we construct a high-quality annotated dataset by combining the module-wise outputs $\mathcal{S}_{I}$ and $\mathcal{S}_{ESG}$ with either $\mathcal{F}_{M}$ or $\mathcal{F}_{E}$, depending on the setting. For each sample, we store the final label and metadata including the reasoning path, reasoning process, supporting rationale, and confidence, ensuring transparency and reproducibility.

\section{Preliminary Analysis of Multi-Agent ESG Reasoning}
Before presenting the full evaluation of \textsc{TRAE}, we conduct a preliminary analysis to examine how different agents reason about ESG relevance assessment, factor labeling, and ambiguity resolution. For each module, we quantitatively analyze the relationship between automation rate and annotation reliability across different routing settings. We perform this analysis independently for each module.

\subsection{Experimental Settings}
\label{sec:ex_settings}
\textsc{TRAE} uses heterogeneous LLM agents to mitigate single-model bias while keeping large-scale annotation cost manageable. Specifically, we use lightweight API models (\texttt{gpt-4.1-mini}, \texttt{deepseek-chat}, and \texttt{gemini-2.5-flash}) as annotation agents, and a stronger Judge LLM (\texttt{gemini-3-flash-preview}), which is invoked only for samples routed to fully automated Ambiguity Resolution. All LLM calls use low-temperature decoding with temperature 0.2, top-$p$ 1.0, and a maximum output length of 3000 tokens; for the debate-based Judge stage, we set \texttt{thinking\_level=high} to strengthen final decision reasoning. For conditional routing, we use $K=3$ agents, resulting in agreement thresholds $\alpha \in \{\frac{1}{3}, \frac{2}{3}, 1\}$, and apply confidence thresholds $\tau \in \{0.90, 0.70, 0.50\}$ to decide whether each sample is automatically finalized or routed to additional verification.

For reliability analysis, we sample 6,000 sentences from \textsc{ESG-Kor}~\cite{lee2024esg}, a Korean ESG information extraction dataset manually labeled by ESG experts, while considering both sector and label balance. We evaluate annotation quality against expert labels using Agreement Accuracy (AA) and Agreement F1-score (AA-F1), and define coverage as the proportion of samples automatically finalized by \textsc{TRAE}. For downstream evaluation, we train BERT-based ESG classifiers on annotated data and report accuracy and F1-score on fixed validation and test sets, each consisting of 1,008 sector- and label-balanced samples.


\subsection{Disagreement in Relevance Assessment and Factor Labeling}
\label{subsec:Disagree}
\begin{figure}[t]
    \centering
    \includegraphics[width=0.90\linewidth]{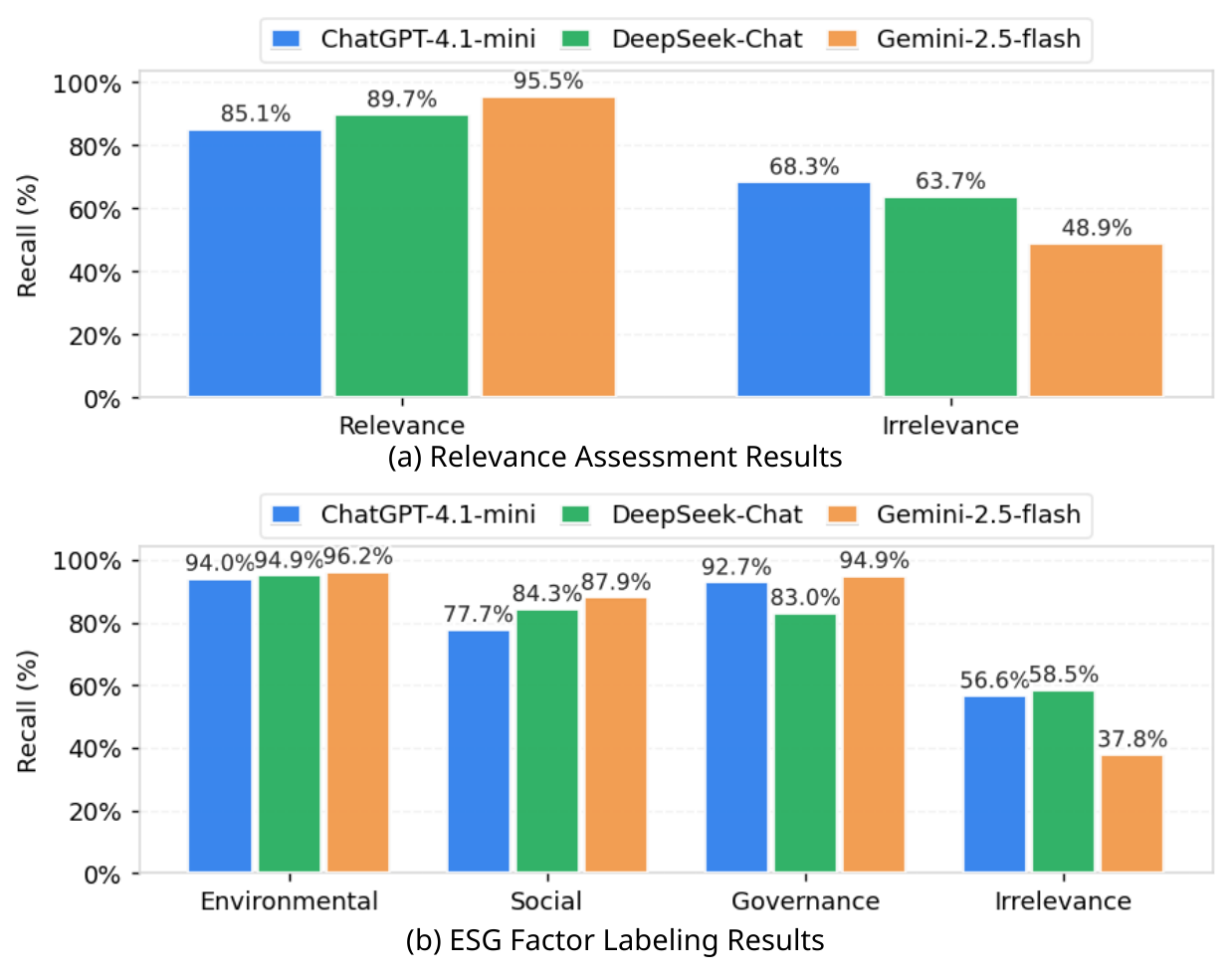}
    \caption{Class-wise recall of single-agent annotations across LLMs.}
    \label{fig3:recall}
\end{figure}

Fig.~\ref{fig3:recall} reports class-wise recall for annotations produced by each individual LLM agent. In both Relevance Assessment and ESG Factor Labeling, we observe noticeable variance across models; in particular, recall for the \texttt{Irrelevant} class is low or shows large inter-model differences. Because models apply different criteria for identifying \texttt{Irrelevant} samples, relying on a single agent can raise stability issues. Inter-agent disagreement is also reflected by Cohen’s $\kappa$ (Table~\ref{tab:kappa_matrix}), which adjusts observed agreement ($p_o$) by chance agreement ($p_e$) and is defined as $\kappa=(p_o-p_e)/(1-p_e)$. For each module, we compute $\kappa$ from the final labels assigned by two models to the same samples (pairwise agreement). Across both modules, pairwise agreement does not reach perfect consistency for all model combinations, indicating that different models can reach different conclusions depending on sample difficulty and context dependence. This supports the view that a single-path annotation approach is insufficient to consistently maintain annotation quality.
\begin{table}[t]
\centering
\tiny
\setlength{\tabcolsep}{1.0pt}
\renewcommand{\arraystretch}{0.8}
\caption{Inter-agent agreement matrices (Cohen's $\kappa \times 100$). Rows follow the same model order in both matrices.}
\label{tab:kappa_matrix}
\resizebox{\columnwidth}{!}{%
\begin{tabular}{lccc @{\hspace{8pt}} ccc}
\toprule
\multicolumn{1}{c}{} & \multicolumn{3}{c}{Relevance Assessment} 
& \multicolumn{3}{c}{ESG Factor Labeling}\\
\cmidrule(lr){2-4}\cmidrule(lr){5-7}
& ChatGPT & DeepSeek & Gemini
& ChatGPT & DeepSeek & Gemini \\
\midrule
ChatGPT  & --    & 77.47 & 62.16  & --    & 84.88 & 82.19 \\
DeepSeek & 77.47 & --    & 69.47  & 84.88 & --    & 80.49 \\
Gemini   & 62.16 & 69.47 & --     & 82.19 & 80.49 & --    \\
\bottomrule
\end{tabular}}
\end{table}

\begin{figure}[h]
    \centering
    \includegraphics[width=\linewidth]{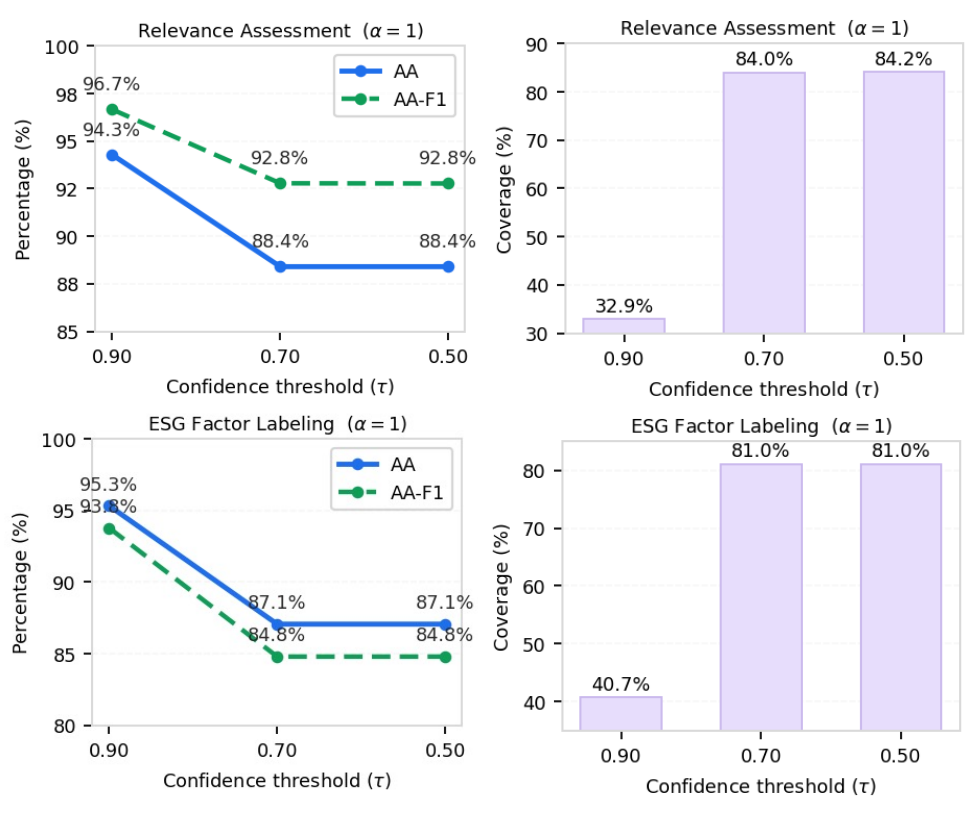}
    \caption{Effect of conditional routing.}
    \label{fig4:eff_routing}
\end{figure}

\subsection{Effect of Conditional Routing}
\label{subsec:Eff_of_route}
\textit{\textbf{Annotation quality.}}
In this experiment, we analyze how conditional routing controls automatic annotation quality and automation coverage by varying the consensus threshold $\alpha_0$ and confidence threshold $\tau_0$. Fig.~\ref{fig4:eff_routing} shows the effect of changing $\tau$ under full consensus ($\alpha=1$). Across both Relevance Assessment and ESG Factor Labeling, a higher confidence threshold yields stronger agreement with expert annotations but lower coverage, while relaxing the threshold substantially increases coverage at the cost of reduced agreement. For example, in Relevance Assessment, $\tau=0.90$ achieves high expert agreement but covers only 32.9\% of samples, whereas $\tau=0.70$ increases coverage to 84.0\%.
These results show that conditional routing provides a direct mechanism for controlling the quality--coverage trade-off. Since higher coverage reduces the number of samples routed to additional reasoning or expert verification, threshold selection directly affects annotation cost as well as quality. Therefore, the routing thresholds should be selected according to the target objective, balancing reliability and annotation efficiency.

\label{subsec:calibration}
\begin{table}[t]
\centering
\scriptsize
\setlength{\tabcolsep}{4pt}
\renewcommand{\arraystretch}{1.05}
\caption{Calibration results for confidence-based routing. Lower ECE and Brier scores indicate better calibration.}
\label{tab:calibration}
\resizebox{0.85\columnwidth}{!}{%
\begin{tabular}{lccc}
\toprule
Setting & AA~(\%) & ECE & Brier \\
\midrule
ChatGPT-4.1-mini  & 80.25 & 0.1376 & 0.1719 \\
DeepSeek-Chat & 80.17 & 0.0943 & 0.1596 \\
Gemini-2.5-flash   & 79.20 & 0.1558 & 0.1816 \\
\midrule
\textsc{TRAE} ($\alpha=1$) 
& 87.06 & 0.0365 & 0.1075 \\
\textsc{TRAE} ($\alpha=1$, $\tau=0.90$) 
& \textbf{95.33} & \textbf{0.0106} & \textbf{0.0442} \\
\bottomrule
\end{tabular}}
\end{table}

\textit{\textbf{Calibration of routing confidence.}}
\textsc{TRAE} uses confidence as a routing signal to determine whether an annotation can be automatically finalized. However, self-reported LLM confidence does not always accurately reflect prediction correctness. We therefore compare raw single-agent confidence and consensus-filtered confidence from a calibration perspective~(Table~\ref{tab:calibration}). We use Expected Calibration Error~(ECE) and Brier score, where ECE measures the gap between predicted confidence and empirical accuracy, and Brier score measures the squared error between confidence and the binary correctness indicator.
The results show that single-agent confidence tends to be overconfident, yielding relatively high ECE and Brier scores. In contrast, applying confidence after full consensus~($\alpha=1$) substantially improves calibration, and combining full consensus with a high confidence threshold~($\tau=0.90$) yields the best-calibrated routing signal. This indicates that consensus-filtered confidence provides a more reliable signal for automatic annotation than raw confidence alone.


\subsection{Effect of Debate on Ambiguity Resolution}
\label{subsec:Eff_of_deb}
\begin{figure}[t!]
    \centering
    \includegraphics[width=\linewidth]{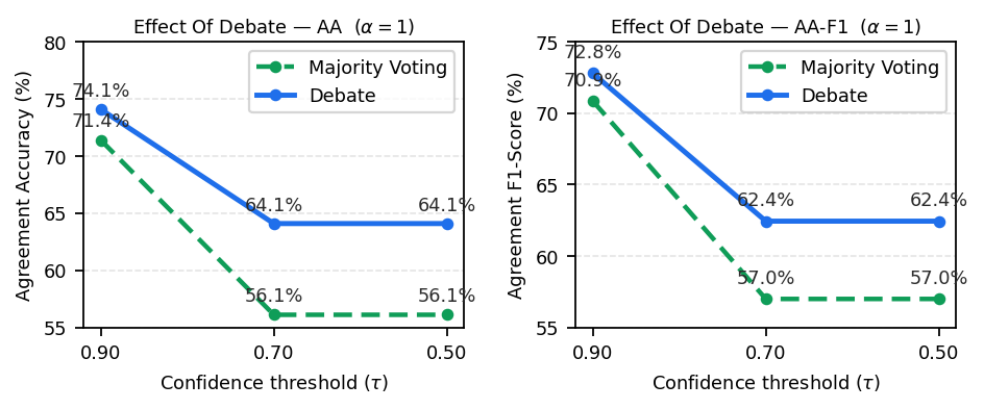}
    \caption{Effect of debate strategy. Comparison between majority voting and debate-based resolution on ambiguous samples}
    \label{fig4:eff_debate}
\end{figure}

We apply either an expert-in-the-loop or a fully automated approach to uncertain samples, depending on the setting. In the fully automated setting, labels are finalized via a debate-based strategy with a Judge LLM. To verify its effectiveness, we test whether debate-based finalization improves annotation quality over majority voting for uncertain samples that do not satisfy the conditional routing criteria. Fig.~\ref{fig4:eff_debate} compares agreement with expert annotations between majority voting and debate within the uncertain region after conditional routing under full consensus ($\alpha=1$) with varying $\tau$.

Results show that debate consistently achieves higher agreement with expert annotations than majority voting across all settings. Whereas majority voting reflects only label frequency, debate elicits strengthened claims and rationales from agents, which the Judge LLM holistically reviews to make the final decision. This suggests that debate yields more trustworthy annotations in uncertain regions by incorporating diverse perspectives.

\begin{table*}[t!]
\centering
\tiny
\setlength{\tabcolsep}{4pt}
\renewcommand{\arraystretch}{0.5}
\caption{\textsc{TRAE} annotation performance under different conditional routing thresholds. $\alpha_{rel},\tau_{rel}$ and $\alpha_{ESG},\tau_{ESG}$ are applied to Relevance Assessment and ESG Factor Labeling, respectively. $|\mathcal{S}_{I}|$, $|\mathcal{S}_{ESG}|$, and $|\mathcal{F}_M|$ denote the numbers of samples finalized at each module. We report coverage and agreement (AA/AA-F1) before debate ($\mathcal{S}_{I}\cup\mathcal{S}_{ESG}$) and after ambiguity resolution ($\mathcal{S}_{I}\cup\mathcal{S}_{ESG}\cup\mathcal{F}_M$).
For readability and space efficiency, we report representative consensus thresholds $\alpha \in \{1,\,\frac{2}{3}\}$, as other threshold settings show only marginal performance differences.}
\label{tab2:threshold_sweep}
\resizebox{0.95\textwidth}{!}{%
\begin{tabular}{cc cc rrr ccc cc}
\toprule
\multirow{2}{*}{$\alpha_{rel}$} & \multirow{2}{*}{$\alpha_{ESG}$}
& \multirow{2}{*}{$\tau_{rel}$} & \multirow{2}{*}{$\tau_{ESG}$}
& \multirow{2}{*}{$|\mathcal{S}_{I}|$} & \multirow{2}{*}{$|\mathcal{S}_{ESG}|$} & \multirow{2}{*}{$|\mathcal{F}_M|$}
& \multicolumn{3}{c}{$\mathcal{S}_{I} \cup \mathcal{S}_{ESG}$}
& \multicolumn{2}{c}{$\mathcal{S}_{I} \cup \mathcal{S}_{ESG} \cup \mathcal{F}_M$} \\
\cmidrule(lr){8-10}\cmidrule(lr){11-12}
& & & & & & & Coverage(\%) & AA(\%) & AA-F1(\%) & AA(\%) & AA-F1(\%) \\
\midrule
\multirow{4}{*}{1} & \multirow{4}{*}{1}
& 0.90 & 0.90 & 254 & 2246 & 3500 & 41.67 & \textbf{94.64} & \textbf{93.09} & \textbf{82.67} & \textbf{81.87} \\
& & 0.90 & 0.70 & 254 & 4657 & 1089 & 81.85 & 86.87 & 84.87 & \underline{82.63} & \underline{81.83} \\
& & 0.70 & 0.90 & 869 & 2021 & 3110 & 48.17 & \underline{91.04} & \underline{88.50} & 82.37 & 81.73 \\
& & 0.70 & 0.70 & 869 & 4285 & 846  & 85.90 & 85.66 & 84.59 & 82.33 & 81.69 \\
\midrule
\multirow{4}{*}{$\frac{2}{3}$} & \multirow{4}{*}{$\frac{2}{3}$}
& 0.90 & 0.90 & 940  & 4559 & 501 & 91.65 & 84.25 & 83.40 & 81.98 & 81.44 \\
& & 0.90 & 0.70 & 940  & 5015 & 45  & 99.25 & 81.73 & 81.11 & 81.45 & 80.87 \\
& & 0.70 & 0.90 & 1344 & 4313 & 343 & 94.28 & 82.29 & 81.92 & 80.78 & 80.58 \\
& & 0.70 & 0.70 & 1344 & 4625 & 31  & 99.48 & 80.63 & 80.36 & 80.43 & 80.19 \\
\bottomrule
\end{tabular}}
\end{table*}

\subsection{Discussion and Implications}
Our preliminary analysis reveals substantial recall variance across individual models and imperfect pairwise $\kappa$ agreement, indicating that single-model annotation is insufficient for consistently maintaining quality and may accumulate errors. We also find that conditional routing controls the trade-off between automation coverage and expert agreement through threshold selection, and that consensus-filtered confidence provides a better-calibrated reliability signal than raw single-agent confidence. These results suggest that conditional routing can serve as a key mechanism for jointly managing annotation cost and quality. Finally, in uncertain regions, debate-based finalization consistently yields higher agreement with experts than majority voting, confirming that identifying uncertain samples and applying follow-up verification stabilizes annotation. Collectively, these observations motivate a trustworthy framework that mitigates agent-level judgment errors, controls the scope of automated annotation, and adds verification for uncertain samples.

\section{Experimental Results}
\label{sec:Exp}
In this section, we evaluate \textsc{TRAE} in terms of annotation reliability, downstream ESG information extraction performance, and practical usability in real-world settings. We use the same models, data, and evaluation metrics as in Section~\ref{sec:ex_settings}. We first assess the fully automated annotation reliability of \textsc{TRAE} against expert labels, compare it with single-agent and multi-agent LLM annotation baselines, and analyze the contribution of each module. We then evaluate the downstream utility of datasets constructed under fully automated and expert-in-the-loop settings. Finally, we apply \textsc{TRAE} to corporate reports from TCFD Hub~\cite{wu2025susgen} to construct \textsc{ESG-En}, a large-scale English ESG information extraction dataset, and validate its downstream utility and practical annotation-cost efficiency.

\subsection{Automatic ESG Annotation Performance}
We evaluate end-to-end \textsc{TRAE} performance under the fully automated setting and quantify the trade-off between annotation quality and automation coverage across conditional-routing thresholds (Table~\ref{tab2:threshold_sweep}). We first compute coverage and agreement with expert annotations for the outputs of Relevance Assessment and ESG Factor Labeling, i.e., $\mathcal{S}_{I}\cup\mathcal{S}_{ESG}$. We then label all remaining uncertain samples via Judge-LLM debate in the Ambiguity Resolution module and evaluate performance after including the resulting dataset $\mathcal{F}_{M}$.

Under high-threshold settings for high-quality annotation, agreement with expert annotations remains very high (94.64\% on $\mathcal{S}_{I}\cup\mathcal{S}_{ESG}$), while coverage is limited to 41.67\%. Relaxing thresholds substantially increases coverage but tends to reduce agreement. At the setting with the highest expert agreement, 41.67\% of the data are automatically finalized before Ambiguity Resolution, maintaining high-quality annotations while substantially reducing the manually annotated portion, potentially lowering cost and improving efficiency. Moreover, after labeling all uncertain samples via the Ambiguity Resolution module, final performance remains relatively stable across threshold settings.

Additionally, following prior work on LLM-based annotators~\cite{xia2024using,Wu2024Enhancing,ni2024afacta}, we compare \textsc{TRAE} with representative annotation settings, including single-agent annotation, self-consistency, and multi-agent aggregation~(Table~\ref{tab:annotation_baseline}). The results show that \textsc{TRAE} achieves higher agreement with expert annotations than all comparison settings. These results suggest that staged ESG reasoning, uncertainty identification, and final verification jointly contribute to the observed improvement over simple multi-agent aggregation. Nevertheless, the fully automated setting still leaves room for improvement, and ambiguous samples require expert verification. In the expert-in-the-loop setting, \textsc{TRAE} achieves approximately 99\% annotation accuracy, supporting the need for a dataset construction strategy that automatically finalizes reliable samples while selectively routing uncertain samples to experts.


\begin{table}[t]
\centering
\footnotesize
\setlength{\tabcolsep}{10pt}
\caption{Comparison with LLM-based annotation settings.}
\label{tab:annotation_baseline}
\begin{tabular}{lrr}
\toprule
Method & AA(\%) & AA-F1(\%) \\
\midrule
Single-Agent & 79.20 & 76.95 \\
Self-Consistency & 79.87 & 78.75 \\
Multi-Agent & 80.22 & 80.18 \\
\textsc{TRAE} (ours) & \textbf{82.67} & \textbf{81.87} \\
\bottomrule
\end{tabular}
\end{table}

\begin{table}[t]
\centering
\small
\setlength{\tabcolsep}{6pt}
\caption{Ablation study of \textsc{TRAE}. A checkmark indicates that the module is enabled.}
\label{tab:ablation}
\resizebox{0.8\linewidth}{!}{%
\begin{tabular}{ccc rr}
\toprule
ESG Factor & Relevance & Ambiguity & \multirow{2}{*}{AA(\%)} & \multirow{2}{*}{AA-F1(\%)} \\
Labeling   & Assessment & Resolution &  &  \\
\midrule
\checkmark &  &  & 80.22 & 80.18 \\
\checkmark & \checkmark &  & 81.18 & 80.59 \\
\checkmark &  & \checkmark & 81.58 & 80.92 \\
\checkmark & \checkmark & \checkmark & \textbf{82.67} & \textbf{81.87} \\
\bottomrule
\end{tabular}}
\end{table}

\subsection{Ablation Study}
We analyze how each \textsc{TRAE} module contributes to annotation quality (Table~\ref{tab:ablation}). Ambiguity Resolution follows the fully automated setting with debate-based finalization. The results show that performance improves when the Relevance Assessment module is added before ESG Factor Labeling and when the Ambiguity Resolution module is applied afterward. This indicates that proactively separating ESG-irrelevant samples contributes to stabilizing overall annotation quality, while the follow-up verification step in Ambiguity Resolution effectively mitigates error accumulation by refining uncertain labels.

\begin{table*}[h]
\centering
\small
\setlength{\tabcolsep}{6pt}
\renewcommand{\arraystretch}{0.1}
\caption{Qualitative case study. The `Judge (summary)' column provides a summarized version of the Judge LLM’s rationale. Keywords in sentences and key cues in the Judge summary are highlighted. Blue indicates correct (Gold=\textsc{TRAE}) and red indicates incorrect (Gold$\neq$\textsc{TRAE}).}
\label{tab:case_cards}
\begin{tabularx}{\textwidth}{p{2.0cm} >{\raggedright\arraybackslash}X >{\raggedright\arraybackslash}X p{1.4cm}}
\toprule
\textbf{Type} & \textbf{Sentence} & \textbf{Judge~(Summary)} & \textbf{Result} \\
\midrule
Label boundary ambiguity
& In addition, the company invested in \kw{Upstream Security} (an overseas mobility innovation firm) to strengthen its \kw{security infrastructure}.
& Interprets \jh{security infrastructure} as a \jh{data security / customer privacy} issue under the Social pillar; treats the Governance ``risk management'' view as less direct for this sentence.
& \ok{Gold:\texttt{S} \textsc{TRAE}: \texttt{S}} \\
\midrule
Keyword-triggered over-inference
& The Autocare System provides vehicle-status information and \kw{eco-driving} information, and notifies replacement/inspection timing for \kw{consumables} (e.g., oil, filters, brake pads), while also checking abnormalities.
& Anchors on \jh{eco-driving} and \jh{consumables management} and infers environmental impact (E), which can \jh{over-extend beyond the dataset policy} for product/function descriptions.
& \bad{Gold:\texttt{I} \textsc{TRAE}: \texttt{E}} \\
\midrule
Policy--definition mismatch
& We will create value for a sustainable future together with all \kw{stakeholders} and share the results in a \kw{balanced} way.
& Applies an implicit \jh{concreteness criterion} (needs concrete actions/practices); thus treats it as a \jh{high-level vision statement} and filters it as \texttt{I}.
& \bad{Gold:\texttt{G} \textsc{TRAE}: \texttt{I}} \\
\bottomrule
\end{tabularx}
\end{table*}

\subsection{Case Study}
We qualitatively analyze annotation cases produced by \textsc{TRAE}. Table~\ref{tab:case_cards} summarizes examples for each case study.

\textit{\textbf{Label boundary ambiguity.}}
ESG annotation can be unstable under a single model’s judgment because the same sentence may map to multiple factors. In the label boundary ambiguity case in Table~\ref{tab:case_cards}, strengthening \emph{security infrastructure} can be interpreted as \texttt{S} (data security/customer privacy) or \texttt{G} (risk management). By prioritizing the \texttt{S} linkage specified in the guideline and finalizing the label as \texttt{S}, the Judge demonstrates that guideline-grounded prioritization can stabilize boundary-case decisions when agents’ judgments diverge.

\textit{\textbf{Keyword-triggered over-inference.}}
Function/product descriptions should be labeled \texttt{I} unless directly linked to policies, programs, outcomes, or management systems, even when ESG-related keywords appear. In the keyword-triggered over-inference case (Table~\ref{tab:case_cards}), the Judge assigns \texttt{E} from cues such as \emph{eco-driving} and \emph{consumables}, although the sentence is essentially a vehicle-management function description. This indicates that keyword-driven interpretation expansion in uncertain regions can cause errors; thus, guidelines should prioritize \texttt{I} when no explicit ESG linkage is stated.

\textit{\textbf{Policy–definition mismatch.}}
In corporate-vision declaratives, annotation errors can arise when the dataset’s operational definitions and the Judge’s implicit criteria are misaligned. In the policy-definition mismatch case (Table~\ref{tab:case_cards}), the Judge labeled the sentence \texttt{I} as a \emph{high-level vision statement} lacking concrete actions; under our policy, however, vision statements that include stakeholder responsibility and responsible-management principles are classified as \texttt{G}. This indicates that automated annotation performance depends on how explicit and consistent the labeling policy is; thus, it is beneficial to predefine sentence-type inclusion/exclusion rules, include boundary-case examples, and use policy-grounded reasoning traces as learning signals.

\subsection{Downstream ESG Information Extraction Performance}
\label{subsec:downstream}
\begin{figure}[t]
    \centering
    \includegraphics[width=\linewidth]{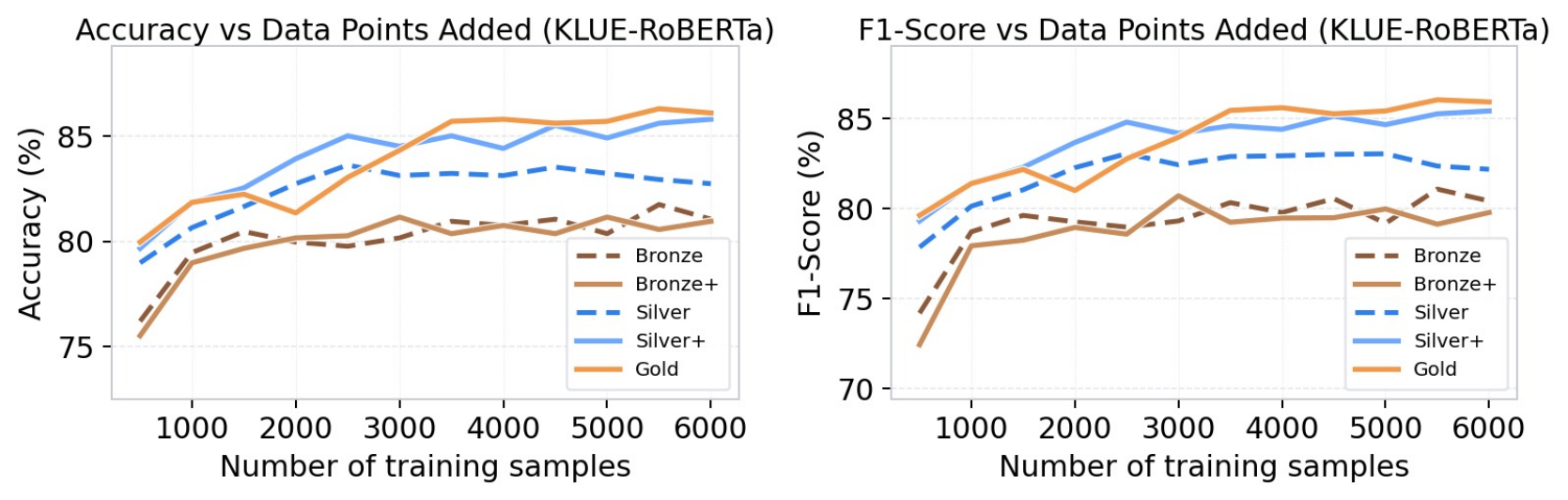}
    \caption{ESG classification performance on \textsc{ESG-Kor} under different dataset construction settings. Test accuracy and F1-score of KLUE-RoBERTa trained with increasing numbers of samples.}
    \label{fig6:downstream}
\end{figure}

As a downstream evaluation, we assess ESG factor classification to examine the effectiveness of the proposed automatic annotation framework. We construct fixed validation and test sets from \textsc{ESG-Kor} by considering sector and label distributions, and train \texttt{KLUE-RoBERTa} classifiers while progressively increasing the amount of \textsc{TRAE}-annotated training data. Based on the two annotation paths in the Ambiguity Resolution module and the threshold settings in Table~\ref{tab2:threshold_sweep}, we define four labeled datasets: \textit{Silver+}, \textit{Silver}, \textit{Bronze+}, and \textit{Bronze}. The \textit{Silver} datasets use an expert-in-the-loop hybrid strategy, whereas the \textit{Bronze} datasets are fully automated. For both strategies, we fix full consensus ($\alpha=1$) and vary the confidence threshold $\tau$: \textit{Silver+}/\textit{Bronze+} use a conservative, quality-focused setting ($\tau=0.90$), while \textit{Silver}/\textit{Bronze} use a more balanced setting ($\tau=0.70$) that considers both quality and cost.
As shown in Fig.~\ref{fig6:downstream}, performance consistently improves as the amount of training data increases across all settings. In particular, \textit{Silver+} approaches the performance of the \textit{Gold} expert-annotated setting, demonstrating that \textsc{TRAE}'s expert-in-the-loop strategy can provide reliable supervision while reducing manual annotation cost.

\begin{figure}[t]
    \centering
    \includegraphics[width=\linewidth]{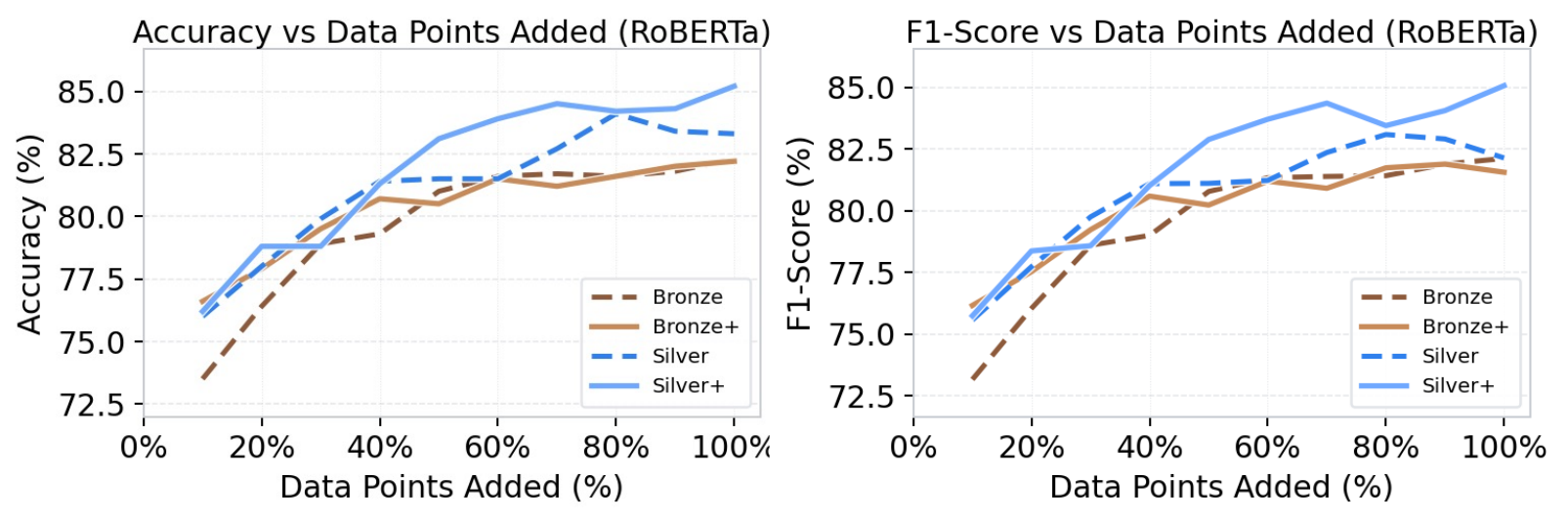}
    \caption{ESG classification performance on \textsc{ESG-En} under different dataset construction settings. Test accuracy and F1-score of RoBERTa trained with increasing numbers of training samples.}
    \label{fig7:ESG-En}
\end{figure}

\begin{figure*}[h]
    \centering
    \includegraphics[width=\linewidth]{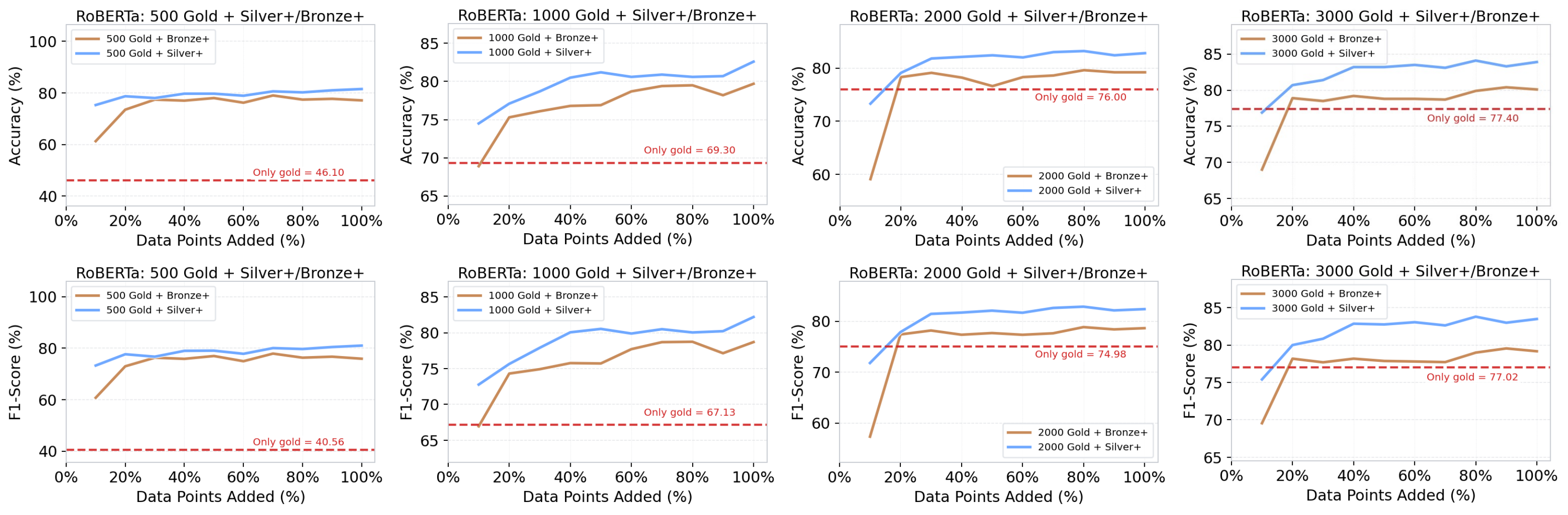}
    \caption{The performance of augmenting a limited number of \textsc{ESG-En}$_{\textit{gold}}$ samples with additional data from \textsc{ESG-En}$_{\textit{silver+}}$ and \textsc{ESG-En}$_{\textit{bronze+}}$. We cap the total training set size at 9,305 samples (the size of the expert-involved set). The red dashed line denotes the performance without augmentation (\textit{Gold}-only).}
    \label{fig8:Limited-ESG-En}
\end{figure*}

\begin{table}[t]
\centering
\small
\setlength{\tabcolsep}{4pt}
\renewcommand{\arraystretch}{1}
\caption{Data statistics for \textsc{ESG-En} dataset construction.}
\label{tab:esgen_source_reports}
\resizebox{\linewidth}{!}{%
\begin{tabular}{l l r r}
\toprule
Company & Industry 
& \makecell[r]{Number of\\Reports} 
& \makecell[r]{Number of\\Sentences} \\
\midrule
AcBel Polytech      & Capital Goods              & 2 & 4{,}795 \\
3M                  & Capital Goods              & 2 & 5{,}824 \\
Ford                & Automobiles and Components & 1 &   851 \\
Toyota              & Automobiles and Components & 1 &   621 \\
Croda International & Chemicals                  & 1 & 4{,}475 \\
SK hynix            & Manufacturing              & 1 &   317 \\
TSMC                & Manufacturing              & 1 &   518 \\
Goldfields          & Metals and Mining          & 3 &   965 \\
Novartis            & Pharmaceutical             & 1 & 1{,}904 \\
\midrule
\multicolumn{2}{l}{Total} & 13 & 20{,}270 \\
\bottomrule
\end{tabular}%
}
\end{table}

\subsection{Expert-in-the-Loop ESG Dataset Construction}
\textit{\textbf{Construction of \textsc{ESG-En.}}}
To assess the practical utility of the proposed \textsc{TRAE}, we construct and validate a new English ESG information extraction and classification dataset, \textsc{ESG-En}, from ESG reports provided by TCFD Hub. Although TCFD Hub covers a wide range of industries, we focus on manufacturing-related companies to ensure within-sector comparability while maintaining company diversity. We use 13 reports from 9 companies and obtain 20,270 sentence samples after preprocessing, including sentence segmentation, duplicate removal, and filtering of noisy sentences. Table~\ref{tab:esgen_source_reports} summarizes the source reports used for \textsc{ESG-En} construction.

To prevent data leakage and enable reliable evaluation, domain experts first label 1,000 samples each for validation and test before the annotation of training data begins. Two ESG domain experts participated in the annotation process, with one serving as the primary annotator and the other supervising guideline consistency and ambiguous-case resolution. Both experts have experience in ESG disclosure analysis, sustainability reporting, and finance-related document interpretation, and used a shared guideline to ensure consistent label decisions. The remaining 18,270 sentences are annotated collaboratively by \textsc{TRAE} and ESG experts. Specifically, experts intervene only for ambiguous samples identified via conditional routing and finalize labels using the same annotation guidelines as \textsc{TRAE}, while referring to model-provided rationales and confidence information. We observe that the performance gap between the fully automated and expert-in-the-loop settings tends to be relatively smaller than in \textsc{ESG-Kor}, which was annotated by experts without model assistance. This suggests that the collaborative setting helps align decision criteria for sentence types with ambiguous label boundaries.

Through this process, we construct \textsc{ESG-En}, an English ESG dataset consisting of 18,270 training samples and 1,000 expert-labeled samples each for validation and test. The dataset is comparable in scale to the single-sector subset of the existing high-quality Korean dataset \textsc{ESG-Kor}, while providing a new English resource for ESG information extraction and classification.

\textit{\textbf{Downstream evaluation on \textsc{ESG-En}.}}
Following the Silver/Bronze construction settings defined in Section~\ref{subsec:downstream}, we construct four \textsc{ESG-En} training variants and evaluate their downstream utility for ESG classification. Table~\ref{tab:esgen_label_dist} reports the label distribution of each variant used in the experiments.

\begin{table}[t!]
\centering
\scriptsize
\setlength{\tabcolsep}{10pt}
\renewcommand{\arraystretch}{0.9}
\caption{Number of samples per label for each dataset in the \textsc{ESG-En} downstream experiments.}
\label{tab:esgen_label_dist}
\resizebox{0.9\columnwidth}{!}{%
\begin{tabular}{lrrrr}
\toprule
Dataset & E & S & G & I \\
\midrule
\textsc{ESG-En}$_{\textit{bronze}}$   & 5,593 & 4,592 & 3,081 & 5,004 \\
\textsc{ESG-En}$_{\textit{bronze+}}$  & 5,601 & 4,750 & 3,456 & 4,463 \\
\textsc{ESG-En}$_{\textit{silver}}$   & 5,484 & 4,581 & 3,104 & 5,101 \\
\textsc{ESG-En}$_{\textit{silver+}}$  & 5,239 & 4,602 & 3,419 & 5,010 \\
\bottomrule
\end{tabular}}
\end{table}

\begin{table}[h!]
\centering
\small
\setlength{\tabcolsep}{5.2pt}
\renewcommand{\arraystretch}{1.15}
\caption{Downstream ESG factor classification performance on \textsc{ESG-En} with training sets constructed under different \textsc{TRAE} annotation settings induced by conditional-routing thresholds.}
\label{tab:esgen_downstream_full}
\resizebox{\columnwidth}{!}{%
\begin{tabular}{l*{10}{r}}
\toprule
\multirow{2}{*}{Model}
& \multicolumn{2}{c}{\textsc{ESG-En}$_{\textit{bronze}}$ }
& \multicolumn{2}{c}{\textsc{ESG-En}$_{\textit{bronze+}}$ }
& \multicolumn{2}{c}{\textsc{ESG-En}$_{\textit{silver}}$ }
& \multicolumn{2}{c}{\textsc{ESG-En}$_{\textit{silver+}}$ }\\
\cmidrule(lr){2-3}\cmidrule(lr){4-5}\cmidrule(lr){6-7}\cmidrule(lr){8-9}
& Acc (\%) & F1 (\%) & Acc (\%) & F1 (\%) & Acc (\%) & F1 (\%) & Acc (\%) & F1 (\%)\\
\midrule
DistilBERT  & 82.20 & 81.82 & 82.60 & 81.69 & 84.30 & 84.07 & 85.20 & 85.03 \\
BERT        & 82.10 & 81.63 & 82.50 & 82.25 & 83.90 & 83.58 & 85.40 & 85.29 \\
ALBERT     & 81.90 & 81.51 & 82.10 & 81.81 & 84.10 & 83.79 & 84.30 & 84.11\\
RoBERTa  & 82.30 & 82.12 & 82.20 & 81.56 & 84.10 & 83.09 & 85.20 & 85.07 \\
\bottomrule
\end{tabular}%
}
\end{table}
Based on these datasets, we vary the amount of training data and examine whether \textsc{TRAE}-annotated data provide reliable supervision for downstream ESG classification~(Fig.~\ref{fig7:ESG-En}). The results show that performance improves steadily as the dataset size increases, indicating that \textsc{TRAE}-annotated data provide effective supervision for model training. At the maximum training size, \textsc{ESG-En}$_{\textit{silver+}}$ achieves 85.20\% accuracy, while \textsc{ESG-En}$_{\textit{silver}}$ achieves 84.10\% accuracy, confirming the meaningful reliability of the constructed dataset.
In addition, Table~\ref{tab:esgen_downstream_full} reports downstream ESG classification performance on \textsc{ESG-En} using DistilBERT, BERT, ALBERT, and RoBERTa, following the classifier settings of prior ESG classification work~\cite{lee2024esg}. The results show that models stably learn ESG information across all dataset settings, with similar performance trends observed across backbones. In particular, \textsc{ESG-En}$_{\textit{silver+}}$, derived from the conservative routing setting, generally achieves the strongest performance. This aligns with our preliminary finding that conservative conditional routing yields more reliable annotations and further supports the reliability of the constructed \textsc{ESG-En} dataset.

Fig.~\ref{fig8:Limited-ESG-En} shows the effect of augmenting limited \textit{Gold} labels with \textsc{TRAE}-annotated data under different levels of label scarcity. We evaluate four \textit{Gold} settings with 500, 1,000, 2,000, and 3,000 expert-labeled samples, and report both accuracy and F1-score. Across all settings, adding \textsc{TRAE}-annotated data consistently improves performance over the \textit{Gold}-only baseline, with particularly large gains in the most label-scarce setting. These results indicate that \textsc{TRAE} provides cost-effective and reliable annotation support and can substantially improve downstream performance even when \textit{Gold} labels are limited.

\section{Practical Cost and Reproducibility}
\label{sec:cost_repro}
\textit{\textbf{Practical cost analysis.}}
To evaluate the practical annotation cost efficiency of \textsc{TRAE}, we compare fully manual labeling with the expert-in-the-loop annotation strategy during \textsc{ESG-En} construction. To avoid overestimating the cost reduction, we use a conservative cost assumption based on general annotator wages rather than ESG expert rates; thus, the estimated fully manual annotation cost can be viewed as a lower bound of the actual expert-labeling cost. Under this assumption, manually labeling approximately 20,000 samples would require about 7--8 weeks of work and cost approximately \$511.
In contrast, \textsc{TRAE} substantially reduces the required expert workload by automatically finalizing reliable samples and routing only highly uncertain samples for expert review. Specifically, in the \textit{silver+} setting, only 49\% of the training data require expert review, reducing the annotation time to about three weeks and the cost to approximately \$236.6. In the \textit{silver} setting, the expert-involvement rate decreases to 12.7\%, requiring only about five days of work and approximately \$81.3. As a result, \textsc{TRAE} reduces annotation cost by 54\% and 84\% compared with fully manual labeling, respectively. In terms of the number of samples requiring direct expert review, it reduces the manual review burden by 51\%--87.3\%, demonstrating its scalability as an annotation strategy for large-scale ESG data construction. The total LLM API inference cost for fully automated annotation of the complete dataset, without expert intervention, was approximately \$106.78.

These results show that \textsc{TRAE} is not intended to fully replace expert judgment with complete automation. Rather, it can be used as a cost-efficient hybrid annotation strategy that automatically finalizes reliable samples while concentrating expert verification on samples with ambiguous label boundaries. For ESG annotation, where label decisions often depend on domain knowledge and policy-specific interpretation, \textsc{TRAE} provides a more practical balance between annotation quality and cost than fully manual expert labeling by selectively reducing the scope of expert intervention.



\textit{\textbf{Reproducibility and artifact release.}}
To improve the reproducibility and verifiability of \textsc{TRAE}, we will provide a public artifact package. The package includes the source code for the end-to-end annotation pipeline, covering Relevance Assessment, ESG Factor Labeling, Ambiguity Resolution, and dataset export. It also provides the prompt templates, labeling guidelines, debate-agent and Judge prompts, routing thresholds, consensus settings, decoding parameters, and execution environment details used at each stage. In addition, we include stage-wise intermediate outputs, such as agent responses, confidence scores, claims and rationales, and routing/reasoning paths, to enable tracing and verification of annotation decisions. Finally, we will release the original data, dataset variants generated under the \textit{Silver+}/\textit{Silver}/\textit{Bronze+}/\textit{Bronze} settings, and the final exported \textsc{ESG-En} dataset with stable identifiers that distinguish LLM-generated annotations from expert annotations, allowing future studies to reproduce and compare results under the same settings. In the current submission, considering the scope of materials available for public release, we first provide prompt templates, labeling guidelines, agent response samples, and dataset samples through a public repository.


\section{Conclusion}
In this study, we propose \textsc{TRAE}, a multi-agent framework for trustworthy automated ESG annotation. \textsc{TRAE} combines multi-stage modules with consensus- and confidence-based conditional routing, automatically finalizing reliable samples while directing uncertain samples to expert verification, thereby achieving both annotation quality and efficiency. Experimental results show that \textsc{TRAE} attains high agreement with expert annotations and outperforms existing automatic annotation approaches. Moreover, ESG information classification models trained on \textsc{TRAE}-annotated data achieve performance close to models trained on \textit{Gold} data, confirming the utility of the automatically annotated data. Furthermore, through collaboration between \textsc{TRAE} and ESG experts, we construct a new dataset, \textsc{ESG-En}, comprising approximately 20,000 instances, and demonstrate that high-quality data can be obtained while substantially reducing annotation cost. 

Recent advances in LLM-based data generation and annotation have driven rapid adoption; however, unclear labeling policies, limited reporting of when and where human intervention occurs, and unreleased intermediate artifacts can undermine reproducibility and trustworthiness, hindering comparative validation under identical settings and real-world deployment. For LLM-involved datasets in particular, without clear documentation of the generation procedure, it is difficult to assess and compare real-world representativeness and potential biases. Therefore, such datasets should provide final labels with a clear distinction between synthetic and real data, explicit reporting of human-intervention stages, and stage-wise reasoning outputs. In this work, we release the reasoning paths and intermediate artifacts produced during dataset construction alongside the dataset, ensuring transparency and strengthening research ethics and reproducibility.

Finally, ESG information extraction and classification is a key prerequisite for sustainability applications such as disclosure analysis, risk assessment, and investment decision support, yet data quality and construction cost remain major bottlenecks for real-world deployment. Going beyond a framework proposal, this work presents expert-collaborative dataset construction and downstream validation, offering a practical pathway to mitigate data scarcity. The resources released in this study are broadly applicable to both industry and research and can complement existing resources across diverse scenarios. Overall, our results expand the foundation of ESG research, alleviate high-cost annotation burdens, and improve the feasibility and trustworthiness of sustainability analysis, supporting stakeholders’ sustainable decision-making and practical societal impact.

\section*{Acknowledgment}
This work was supported by the Institute of Information \& Communications Technology Planning \& Evaluation (IITP) grant funded by the Korea government (MSIT) (No. RS-2020-II201373, Artificial Intelligence Graduate School Program (Hanyang University)), the Korea Planning \& Evaluation Institute of Industrial Technology (KEIT), funded by the Ministry of Trade, Industry and Energy (MOTIE), Korea (No. RS-2026-25509027, Development and Demonstration of an Ontology- and On-Device AI-Based Autonomous Operation Agent System for Semiconductor Wafer Manufacturing Processes), and the National Research Foundation of Korea (NRF) grant funded by the Korea government (MSIT) (No. RS-2024-0040780).

\bibliographystyle{IEEEtran}
\bibliography{references}

\end{document}